\documentclass[aps,twocolumn,showpacs,amsmath,prb,amssymb,superscriptaddress]{revtex4-1}

\pdfoutput=1

\usepackage[export]{adjustbox}
\usepackage{amsfonts}
\usepackage{amsmath}
\usepackage[makeroom]{cancel}
\usepackage[english]{babel}
\usepackage[T1]{fontenc}
\usepackage{times}
\usepackage{mathrsfs}
\usepackage{graphicx}
\usepackage{dcolumn}
\usepackage{bm}
\usepackage{wasysym}
\usepackage[colorlinks,bookmarks=true,citecolor=blue,linkcolor=red,urlcolor=blue]{hyperref}
\usepackage[tight, FIGTOPCAP, hang, raggedright, nooneline]{subfigure}
\usepackage{hyperref}
\usepackage{xcolor}
\usepackage{epsfig}
\usepackage{amssymb}
\usepackage{lipsum}
\usepackage{algorithm}
\usepackage{algpseudocode}
\usepackage{comment}

\renewcommand{\vec}[1]{\mathbf{#1}}
\newcommand*\diff{\mathop{}\!\mathrm{d}}

\begin{document}
\title{Driven quantum dot coupled to a fractional quantum Hall edge}
\author{Glenn Wagner}
\affiliation{Rudolf Peierls Centre for Theoretical Physics, Parks Road, Oxford, OX1 3PU, UK}
	\author{Dung X. Nguyen}
	\affiliation{Rudolf Peierls Centre for Theoretical Physics, Parks Road, Oxford, OX1 3PU, UK}
	
	\author{Dmitry L. Kovrizhin}
	\affiliation{Rudolf Peierls Centre for Theoretical Physics, Parks Road, Oxford, OX1 3PU, UK}
	\affiliation{NRC Kurchatov institute, 1 Kurchatov Square, 123182, Moscow, Russia}
	
	\author{Steven H. Simon}
	\affiliation{Rudolf Peierls Centre for Theoretical Physics, Parks Road, Oxford, OX1 3PU, UK}
	
\begin{abstract}
We study a model of a quantum dot coupled to a quantum Hall edge of the Laughlin state, taking into account short-range interactions between the dot and the edge. This system has been studied experimentally in electron quantum optics in the context of single particle sources. We consider driving the dot out of equilibrium by a time-dependent bias voltage. We calculate the resulting current on the edge by applying the Kubo formula to the bosonized Hamiltonian. The Hamiltonian of this system can also be mapped to the spin-boson model and in this picture, the current can be perturbatively calculated using the non-interacting blip approximation (NIBA). We show that both methods of solution are in fact equivalent. We present numerics demonstrating that the perturbative approaches capture the essential physics at early times, although they fail to capture the charge quantization (or lack thereof) in the current pulses integrated over long times.
\end{abstract}

\maketitle

\section{Introduction}
	
\begin{figure}[b]
\centering
\includegraphics[width=0.8\columnwidth]{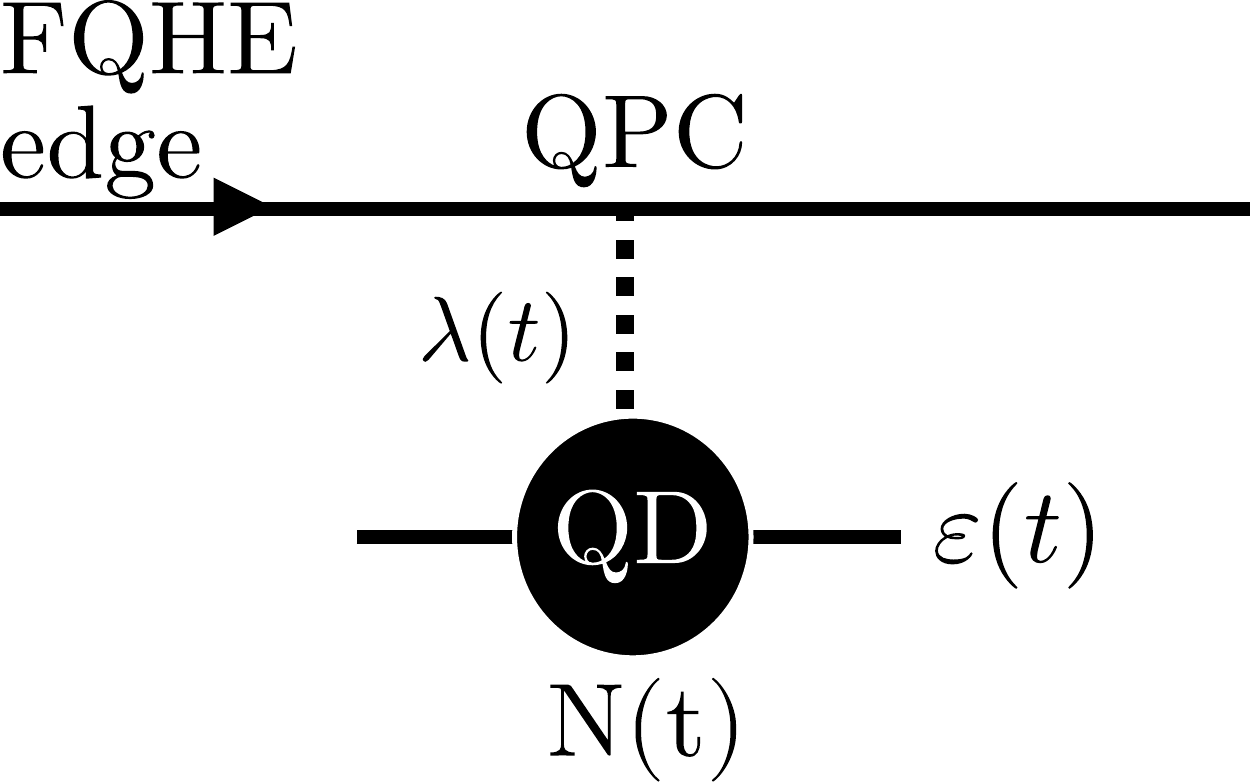}
\caption{Schematic model of the experimental setup showing a quantum dot (QD) coupled to an FQHE edge state. The edge is coupled to the dot via a quantum point contact (QPC) at $x=0$. The gate voltage can be used to tune the tunneling $\lambda(t)$ between the dot and the edge. A single energy level $\varepsilon(t)$ on the quantum dot is controlled via applied bias voltage. We study the occupation number $N(t)$ of the dot over time.}\label{fig:setup}
\end{figure}

Electron quantum optics (EQO) is a field devoted to the study and manipulation of single electron excitations. A promising platform for EQO experiments has proven to be the edge states of the fractional quantum Hall effect (FQHE) \cite{Review3}: Since the edge states are chiral, there is no possibility of backscattering. An important ingredient in any EQO experiment is a single-electron source and in the present paper we study a model of such a device. The model is comprised of a chiral FQHE Laughlin edge state, which we model as a Luttinger liquid. The edge is coupled to a quantum dot via a quantum point contact (QPC). By applying a time-dependent gate voltage to the dot, one is able to obtain current pulses on the edge. The set-up is shown in Fig.~\ref{fig:setup}. In a recent paper \cite{OurLetter}, the present authors highlighted the effect of interactions in this set-up, resulting in a non-quantized charge in the pulses on the edge. In this paper, we use a different set of techniques to study the same problem, which allows us to extend our results to different regimes. 


For these purposes we focus on the mapping, originally proposed by Furusaki and Matveev~\cite{Furusaki2002}, between the spin-boson model, and the chiral Luttinger liquid coupled to a single energy level. The spin-boson model---which describes a spin interacting with a bosonic heat bath---is an important archetype of a quantum dissipative system~\cite{weiss2012quantum,Leggett1987,Orth2013nonperturbative,LeHur2009,LeHur2018,Sassetti1990,Bulla2005,LeHur2005}. Among other applications it has been used to describe decoherence of qubits in quantum information science~\cite{schoelkopf2003qubits,makhlin2003dissipation}. In this paper we present the details of the mapping between the spin-boson model and a fractional quantum Hall (FQHE) edge state coupled to a quantum dot (QD) ~\cite{Furusaki2002}.

 This correspondence proves to be useful for our purposes, since many analytical and numerical techniques have been developed for the spin-boson model including the  non-interacting blip approximation (NIBA) \cite{Grifoni1999}, generalized master equation \cite{Hartmann2000}, stochastic Schr\"odinger equation description \cite{Orth2013nonperturbative}, Bethe-ansatz \cite{CEDRASCHI2001} solution, numerical renormalization group~\cite{Bulla2005,Anders2005}, exact mapping between the spin-boson and the Kondo model \cite{Furusaki2002}, and most recently tensor network methods \cite{Wall2016,Strathearn2018}. Conversely, it is possible to envisage the QD set-up as a quantum simulator for the spin-boson model.

For pedagogical purposes and as a consistency check of the mapping from Luttinger liquid to spin-boson model, we use perturbation theory to calculate the current downstream from the dot within both original and dual descriptions. In the bosonization language, the current is obtained from the Green's functions for the vertex operators, while in the spin-boson language we apply the perturbative solution described in [\onlinecite{Grifoni1999}], the so-called NIBA. Using perturbation theory to the second order in the tunneling between the dot and the edge we show that both pictures yield identical results. We benchmark this perturbative solution against non-perturbative schemes that are valid at special points in the parameter range of the Hamiltonian. For example, in the case where the FQHE edge is in the integer quantum Hall regime, the problem can be solved exactly. Although this solution has been obtained previously, we present a simpler derivation of the integer quantum Hall result. The perturbative solution is shown to be a very good approximation to that solution at early times.  
	

There have been numerous theoretical works~\cite{Yang2013,Galda2011,Rylands2009,Goldstein2010,LeHur2005,Komnik,Wachter2007,Goldstein2011,Goldstein2010a,Goldstein,Goldstein2010b,Fiete2010,Klinovaja,Martin1,Martin2,Martin3,Martin4,Filippone} on single-particle electron emitters. However, in contrast to the previous body of work, here we focus on the non-equilibrium setting, where the energy level of the quantum dot varies with time under applied time-dependent bias voltage. The time evolution of the system in this setup is an interesting problem both from the theoretical and experimental perspective.  

Similar set-ups have also been implemented in recent experiments~\cite{Feve2007,Hermelin2011,McNeil2011,Dubois2013}. In particular, in Ref.~[\onlinecite{Feve2007}] the QD is driven by a square wave voltage. This experiment studies the edge in the integer quantum Hall regime. Current pulses are detected on the edge downstream from the QD. See also reviews \cite{Review1,Review2,Review3} for more details of experiments. We hope that the formulae that we derive will be useful for experimentalists when analysing the measured current pulses. In particular, we have the central result \eqref{eq:electron_current} for the current on the edge, from which an even simpler analytical expression \eqref{eq:analytic_current} can be derived for a particular drive protocol. 

The outline of the paper is the following. In section \ref{sec:Model} we introduce the model and notations. Section \ref{sec:Mapping_to_spin-boson} presents the mapping between the QD-FQHE model and the spin-boson model, see also Ref.~[\onlinecite{OurLetter}]. In section \ref{sec:Current_from_bosonization} we bosonize the original Hamiltonian and obtain a perturbative solution which is valid at short times. In section \ref{sec:Current_from_SB} we use the NIBA solution to derive the expression for the current, and in section \ref{sec:Equivalence_of_solutions} we present a comparison between the results obtained using these approaches. The section \ref{sec:non_perturbative} presents numerical results comparing the perturbative solution to two exact solutions valid in certain parameter regimes. In section \ref{sec:Analytical_limits} we derive analytical results for the current in a number of physically interesting limits.  The exact calculation in the case of integer quantum Hall edge state, and further details are presented in the Appendices.
	
\section{The model and its mapping to the spin-boson problem}
\subsection{Model}\label{sec:Model}
In this paper we study a model of a quantum dot or anti-dot with a single energy level that is coupled to a FQHE edge state. The model is described by the time-dependent Hamiltonian
\begin{equation}\label{eq:Hamiltonian}
\hat H(t)=\hat H_0(t)+\hat H_\textrm{tun}(t)+\hat H_\textrm{int},
\end{equation}
where the terms on the right-hand side correspond to the Hamiltonian of the dot/anti-dot and the edge ($\hat{H}_0$), the tunneling at the QPC ($\hat H_\textrm{tun}$), and the interactions between the dot/anti-dot and the edge ($\hat H_\textrm{int}$) respectively. The Hamiltonian $\hat{H}_0(t)$ describing the dot/anti-dot and the edge without coupling is given by the following expression

\begin{equation}
\label{eq:FreeHam}
\hat{H}_0(t)=\varepsilon(t)\hat{S}_z+ \frac{v}{2}\int \frac{dx}{2\pi}\ (\partial_x \hat\varphi)^2.
\end{equation}

Here the first term on the right-hand side is the energy of a quantum dot/anti-dot with a single level, and we use operators $\hat S^+/\hat S^-$ to describe creation/annihilation of a spinless electron or quasi-electron on this level. A quantum dot is situated on the outside of the quantum Hall fluid and can only host electrons. If we want to host Laughlin quasiparticles we need to consider an anti-dot, which has to be located inside the quantum Hall fluid. Since we want to consider both the cases of electron and quasiparticle tunneling, we need to allow for both the case of a dot and anti-dot. In the case of a dot, $\hat S^+$ creates a electron with charge $q=-e$ (with $e>0$ being the elementary charge), whereas in the case of an anti-dot we assume we have quasi-electrons tunneling so that $\hat S^+$ creates a quasi-electron with charge $q=-\nu e$. Note that the case of holes or quasi-holes is equivalent to electrons or quasi-electrons which tunnel in the opposite direction therefore we do not need to consider this case separately. For conciseness, in the following we will use term "quantum dot" to refer to either a dot or anti-dot, where it is understood that we are considering a dot when electrons are tunneling and an anti-dot when quasi-electrons are tunneling.

While the spin operators describing the level on the dot satisfy commutation relations which are different to those of electron or quasi-electron operators we show in Appendix \ref{sec:App_Klein_factors} that the associated statistical phase does not enter the results for the current and therefore our spin representation is justified for our purposes. The presence or absence of a particle on the dot is measured by the operator $\hat N= \hat{S}_z+1/2$. The energy level of the dot is a function of time $\varepsilon(t)$ and is controlled by an applied time-dependent bias voltage. In comparison with the previous work \cite{Furusaki2002,Galda2011,Goldstein,Goldstein2011,Goldstein2010,Goldstein2010b,Goldstein2010a,Komnik} which focussed on the case of a constant bias $\varepsilon(t)=\textrm{const.}$, here we study a time-dependent problem. 

The second term in equation \eqref{eq:FreeHam} is the bosonized Hamiltonian of an FQHE edge with the length $L$ describing a Laughlin state at filling fraction $\nu=1/(2n+1)$ where $n=0,1,2,\dots$\cite{Wen1992}. The bosonic field can be expanded in its eigenmodes with momentum $k=2\pi m/L$, $m\in Z$ as follows \cite{vonDelft1998}
\begin{equation}
\hat{\varphi}(x)=-\sum_{k>0}\sqrt{\frac{2\pi}{k L}}(\hat b_k e^{i k x}
+\hat{b}^{\dagger}_k e^{-ikx}) e^{-ka/2},
\label{eq:eigenmode_expansion}
\end{equation}
where $a$ is the short-distance cutoff. The commutation relations of the bosonic operators $\hat{b}_k$ are given by $[\hat{b}_k,\hat{b}^\dagger_{k'}]=\delta_{kk'}.$

The electron and the quasiparticle operators in the bosonization language are vertex operators of the form \cite{Wen1992}
\begin{equation}
\label{eq:fermion}
\hat{\psi}(x)=\frac{1}{\sqrt{2\pi}}\bigg(\frac{L}{2\pi}\bigg)^{-\frac{\gamma^2}{2}}:\!e^{-i\gamma\hat\varphi(x)}\!:,
\end{equation} 
where $\gamma=1/\sqrt{\nu}$ or $\gamma=\sqrt{\nu}$ for electrons with charge $-e$ and quasiparticles with charge $-\nu e$ correspondingly, and $:\!\cdots\!:$ denotes normal ordering. Note that we have omitted Klein factors since in our problem they do not affect the results for the current, see Appendix \ref{sec:App_Klein_factors}. Using results of Ref.~[\onlinecite{vonDelft1998}] we rewrite the expression for the vertex operators as
\begin{equation}\label{eq:fermion1}
\hat{\psi}(x)=\frac{1}{\sqrt{2\pi}}a^{-\frac{\gamma^2}{2}}e^{-i\gamma\hat\varphi(x)}.
\end{equation} 
The Hamiltonian describing tunneling of electrons or quasiparticles between the dot and the edge is given by 
\begin{equation}
\label{eq:coupling}
\hat{H}_{\mathrm{tun}}(t)=\lambda(t)\hat\psi^\dagger(0) \hat S^-+h.c,
\end{equation}
where the tunnelling amplitude $\lambda(t)$ is any time-dependent function. Below we will focus on the the specific case of $\lambda(t)=\lambda \theta(t)$, which corresponds to the situation when the tunneling has been suddenly turned on at $t=0$. We have also studied the effects of a gradual switching of the tunneling in the form $\lambda(t)=\lambda\tanh(t/t_s)$, where $t_s$ is some time-scale. However, we find that it does not change the qualitative behaviour of the current. 
	
We model the Coulomb interactions between the dot and the edge using the following Hamiltonian
\begin{equation}
\hat{H}_{\mathrm{int}}=-\gamma\frac{g}{2\pi}\partial_x\hat\varphi(0)\hat{S}_z,
\label{eq:capacitive_coupling}
\end{equation}
where we used a bosonized form of the charge density operator on the edge $\hat{\rho}(x)=+e\sqrt{\nu}\partial_x\hat\varphi/2\pi$, and $g>0$ is the interaction strength. A detailed discussion of the effects of Coulomb interactions has been presented in our previous work \cite{OurLetter}, where we showed that the interactions of the form (\ref{eq:capacitive_coupling}) amount to rescaling of the interaction constant $\gamma$ such that
\begin{equation}
\tilde{\gamma}=\gamma\left(1-\frac{g}{2\pi v}\right).
\label{eq:interactions}
\end{equation}
By performing the unitary transformation discussed in subsection \ref{sec:Mapping_to_spin-boson} below, it can be seen that the Coulomb interaction term can be eliminated by this rescaling.
The equilibrium occupation of the quantum dot with the Hamiltonian \eqref{eq:Hamiltonian} has been investigated perturbatively in [\onlinecite{Furusaki2002}], where the authors found two regimes depending on the strength of $\tilde\gamma$. In the weak-tunneling limit if $\tilde\gamma>\sqrt{1/2}$  there is a discontinuity in the occupation number at $\varepsilon=0$, whereas the latter is continuous for $\tilde\gamma<\sqrt{1/2}$. 

\subsection{Transformation to the spin-boson problem}
\label{sec:Mapping_to_spin-boson}
In this section we will study the dynamics of the dot-edge system under the Hamiltonian \eqref{eq:Hamiltonian} using a mapping to the spin-boson model. This mapping is performed using an unitary transformation introduced in Ref.~[\onlinecite{Furusaki2002}]. Let us define a unitary operator $\hat{U}_1=\exp[-i\gamma\hat\varphi(0)\hat S_z]$. Using the transformation $\hat{H}_1=\hat{U_1}^\dagger \hat H \hat{U_1}$ and expressing the bosonic fields in terms of their modes via Eq.~\eqref{eq:eigenmode_expansion} we obtain
\begin{equation}
\label{eq:Hamsb}
\hat{H}_1=\varepsilon(t)\hat S_z + \Delta(t) \hat S_x +\sum_{k>0}\omega_k\hat b^\dagger_k \hat b_k-i\hat S_z\sum_{k>0}\eta_k (\hat  b_k -\hat b^\dagger_k),
\end{equation}
where $\omega_k=v k$ and
\begin{equation}
\Delta(t)=\lambda(t)\sqrt{\frac{2}{\pi}}a^{-\frac{\gamma^2}{2}},\ \  \eta_k=v\tilde\gamma\sqrt{\frac{2\pi k}{L}}e^{-ka/2}.
\label{eq:Delta_Lambda_def}
\end{equation}
The Hamiltonian in Eq. (\ref{eq:Hamsb}) has the standard spin-boson form\cite{Leggett1987}. The first two terms of equation \eqref{eq:Hamsb} represent a spin-1/2 degree of freedom coupled to a time-dependent magnetic field  $\vec{B}(t)=\varepsilon(t)\hat{\vec{z}}+\Delta(t)\hat{\vec{x}}$. The last two terms describe the bosonic bath, and the coupling of the spin to the bath respectively. The spectral function of the bosonic bath is given by the expression
\begin{equation}
\label{eq:spectralJ}
J(\omega)=\pi\sum_{k>0} \eta_k^2 \delta(\omega-\omega_k)=2\pi\alpha\omega\theta(\omega) e^{-a\omega/v}.
\end{equation}
This spectral function corresponds to the spin-boson model with Ohmic dissipation and with a dimensionless coupling constant $\alpha=\tilde\gamma^2/2$. 

As the next step we refermionize the Hamiltonian \eqref{eq:Hamsb} using another unitary transformation $\hat{U}_2=\exp[i\tilde\gamma\hat\varphi(0)\hat S_z]$ such that $\hat{H}_2=\hat{U_2}^\dagger \hat H_1 \hat{U_2}$ arriving
at the Hamiltonian which has the same form as \eqref{eq:Hamiltonian} but without the interaction term, and with renormalized tunnelling strength
\begin{equation}
\hat{H}_{2,\textrm{tun}}=\tilde\lambda(t)\hat{\tilde\psi}^\dagger(0) \hat S^-+h.c.,
\end{equation}
where the tunneling is given by $\tilde\lambda(t)=a^{(\tilde\gamma^2-\gamma^2)/2}\lambda(t)$,
and the corresponding $\Delta(t)=\tilde\lambda(t)\sqrt{2/\pi}a^{-\frac{\tilde\gamma^2}{2}}$. After refermionization the vertex operators assume the following form
\begin{equation}
\hat{\tilde\psi}(x)=\frac{1}{\sqrt{2\pi}}a^{-\frac{\tilde\gamma^2}{2}}e^{-i\tilde\gamma\hat\varphi(x)}.
\end{equation}

The mapping between the quantum dot system and the spin-boson model is useful, since the latter has been a well-studied problem. It is an archetype of an open quantum system and as such, many numerical techniques have been developed for it. 
On the other hand, the spin-boson model is difficult to model experimentally. Suggested experiments include trapped ions\cite{Porras2008,Lemmer2018} and superconducting circuits \cite{Lappakangas2018,Magazzu2018}. The quantum Hall edge set-up discussed in this work is an alternative experimental proposal to study the spin-boson model in the sense of a quantum simulator. Since we know the Hamiltonian for the quantum Hall edge set-up, one can perform the experiment in order to learn about the spin-boson model in regimes that are difficult to access numerically.  
	\\	
	\!\!\!\!\!\!\!\!\!\!\!	\begin{table}[ht]
		\caption{Dictionary of the spin-boson mapping}
		\begin{tabular}{| c | c |}
			\hline
			
			\textbf{QD + FQHE edge}  &  \textbf{Spin-boson model} \\
			
			\hline
			\hline             
			occupation of the QD $\hat N$ &spin $\hat S_z=\hat N-1/2$ \\
			bosonic operators $\hat b_k$ & heat bath bosons $\hat b_k$ \\
			vertex exponent $\tilde\gamma$ & spin-bath coupling $\alpha=\tilde\gamma^2/2$ \\
			QD voltage bias $\varepsilon(t)$ & magnetic field $B_z=\varepsilon(t)$ \\
			tunneling $\lambda(t)$ & magnetic field $B_x=\Delta(t)$ \\
			IQHE case $\tilde\gamma=1$ & Toulouse limit $\alpha=1/2$ \\
			
			\hline
			
		\end{tabular}
		\label{table:nonlin}
	\end{table}

\section{Calculation of the current}
In this section we derive the expression for the time dependent current using two different perturbative approaches. The first calculation in done in the bosonized Luttinger liquid picture, whereas the second calculation uses well-known results from the spin-boson model. Both give the same answer, which provides a consistency check of the mapping discussed above.
\subsection{Perturbation theory approach for the bosonized Hamiltonian}
\label{sec:Current_from_bosonization}
Here we assume that the quantum dot is weakly-coupled (by tunneling) to the FQH edge, and we work in the interaction representation where the tunneling Hamiltonian plays the role of the interactions. The current operator is given by
\begin{equation}
\label{eq:eomI}
\hat{I}(t)=-\tilde q\frac{\diff \hat N}{\diff t}=-i\tilde q[\hat H_{\textrm{2,tun}}(t),\hat N(t)],
\end{equation}
where $\hat N(t)$ is the number operator on the quantum dot. We note that, as we have shown in our previous work~\cite{OurLetter}, one has to take into account the renormalization of the charge
\begin{equation}
\tilde{q}=q\left(1-\frac{g}{2\pi v}\right).
\label{eq:interactions}
\end{equation}
Intuitively, this renormalization accounts for the fact that the charge density on the edge will be depleted close to the QPC due to Coulomb repulsion.
In order to calculate the expectation of the current to leading order in perturbation theory, we use the Kubo formula
\begin{equation}
\label{eq:CurrentPer}
I(t)=-i\int_{0}^t\langle [\hat{I}(t),\hat H_{\textrm{2,tun}}(t')]\rangle \diff t',
\end{equation}
where we have assumed that the perturbation switches on at $t=0$. Using this expression we arrive at the result for the time-dependent current
\begin{multline}
\label{current4}
I(t)=-\tilde q\int_{0}^t d t'\tilde\lambda(t)\tilde\lambda(t')\\ \times\bigg(e^{i\Omega(t')-i\Omega(t)}[(1-n_a)\Phi(t-t')-n_a\Phi(t'-t)]\\+e^{i\Omega(t)-i\Omega(t')}[(1-n_a)\Phi(t'-t)-n_a\Phi(t-t')]\bigg)
\end{multline}
where we defined $\Omega(t)=\int_0^t d s\ \varepsilon(s) $, and $n_a=\langle\hat  N(0)\rangle$ is the initial occupation of the quantum dot. The correlation function $\Phi(\tau)=\langle \hat{\tilde\psi}(0,\tau) \hat{\tilde\psi}^\dagger(0,0)\rangle$ is given by the expression
\begin{equation}
\Phi(\tau)=\frac{1}{2\pi}[iv\tau_B\sinh(\tau/\tau_B-ia/(v\tau_B))]^{-\tilde\gamma^2},
\label{eq:propagator}
\end{equation}  
where we have introduced a characteristic timescale $\tau_B=\beta/\pi$ associated with temperature $T=1/k_B\beta$, see details of the derivation of the expression for the current in Appendix \ref{sec:perturb}. This perturbative expression for the current represents one of the central results of our work.


As usual in the case of the Kubo formula, we have obtained an early time result. The expression of the current expectation value \eqref{current4} allows us to calculate the current profile at short times. In the perturbative calculation, we assume that during time $t$ the change in the quantum dot occupation number is small. At high temperatures the characteristic time scale is $t\ll\frac{1}{\lambda}(v\tau_B)^{\tilde\gamma^2/2}$, whereas at low temperatures it is given by $t\ll (\tilde\lambda^{-1}v^{\tilde\gamma^2/2})^{\frac{1}{1-\tilde\gamma^2/2}}$. The only dependence on the coupling $\tilde\lambda$ in the perturbative solution is an overall prefactor $\tilde\lambda^2$. Thus the shape of the perturbative current is independent of $\tilde\lambda$, which corresponds to the limit of small $\tilde\lambda$. Therefore, this approach cannot be used to calculate the change in an entire current pulse, since this requires integrating over all times. However, based on the arguments in Ref.~[\onlinecite{OurLetter}], we know that this charge will be less than the electron or quasiparticle charge due to the interaction effects.

\subsection{Non-interacting blip approximation for the spin-boson model}\label{sec:Current_from_SB}
	
Using the Feynman-Vernon influence functional, the authors of [\onlinecite{Grifoni1999}] presented a path-integral solution for the time-evolution of the reduced density matrix of the two-level system  $\rho_{\sigma\sigma'}(t)$ by integrating out exactly the heat bath degrees of freedom.  Using this density matrix one can obtain the occupation of the dot $N(t)$ as well as the current profile $I(t)$. For a general initial condition $\rho_{\sigma_0\sigma_0'}(t_0)$ the time evolution of the density matrix is given by the equation
\begin{equation}\label{exact_FV}
\rho_{\sigma\sigma'}(t)=\int\mathcal{D}\sigma\mathcal{D}\sigma'\mathcal{A}[\sigma]\mathcal{A}^*[\sigma']\mathcal{F}[\sigma,\sigma']\rho_{\sigma_0\sigma_0'}(t_0)
\end{equation}
where the integral is taken over all possible spin paths $\sigma(t)$. Here  $\mathcal{A}[\sigma]$ is the probability amplitude for the system to follow the path $\sigma(t)$ in the absence of heat-bath fluctuations, and $\mathcal{F}[\sigma,\sigma']$ is the Feynman-Vernon influence functional\cite{feynman2000theory,Grifoni1999} which takes into account the heat bath, see Ref. [\onlinecite{Grifoni1999}]. The equation (\ref{exact_FV}) is exact, however in order to use it one has to evaluate the path integral over all possible spin paths. In practice the path integral is turned into a sum over spin flips and we integrate over all possible times at which spin flips occur. In our numerical calculations this series is truncated at a some fixed number of spin-flips.
	
The initial condition for this procedure corresponding to the dot having initial occupation $n_a$ is given by $\langle\sigma_z\rangle(t=0)=2n_a-1$. Assuming the spin subsystem evolves from a pure state, it is shown in \cite{Grifoni1999} that the time evolution reads
\begin{equation}
\langle\sigma_z(t)\rangle=(2n_a-1)P_1^{(s)}(t)+P_1^{(a)}(t),
\label{eq:sigma_z}
\end{equation}
where $P_1^{(s)}(t)$ and $P_1^{(a)}(t)$ are obtained from the series expansion in $\Delta$. Each factor of $\Delta$ includes an additional time integral, hence limiting the maximum order of perturbation theory which we can evaluate numerically. Up to the second order in $\Delta$ we have\footnote{Our Hamiltonian differs from that in [\onlinecite{Grifoni1999}] by a minus sign in the definition of $\varepsilon(t)$ and $\Delta(t)$, hence the expressions below differ in the sign of $\Omega(t)$ from the results in that paper.}
\begin{align}\nonumber
P_1^{(s)}(t)&=1-\int_0^t d t_2\int_0^{t_2}d t_1 \Delta(t_2)\Delta(t_1) \cos(\Omega(t_1)-\Omega(t_2))\\ &\times e^{-Q'(t_2-t_1)}\cos(Q''(t_2-t_1)+Q''(t_1)-Q''(t_2)),\nonumber
\end{align}
and
\begin{align}\nonumber
P_1^{(a)}(t)&=\int_0^t d t_2\int_0^{t_2} d t_1 \Delta(t_2)\Delta(t_1) \sin(\Omega(t_1)-\Omega(t_2))\\ &\times e^{-Q'(t_2-t_1)}\sin(Q''(t_2-t_1)+Q''(t_1)-Q''(t_2)).\nonumber
\end{align}
The expansion to second order in $\Delta$ means that we consider paths with at most two spin flips. In the context of the spin-boson model this truncation is called a non-interacting blip approximation (NIBA). For an Ohmic heat bath with spectral function \eqref{eq:spectralJ}, the exact expressions for the functions $Q'(\tau)$ and $Q''(\tau)$ are given in [\onlinecite{Grifoni1997}], and which in the limit of small cutoff $a$ read
\begin{align}
Q'(\tau)&=\alpha\ln(1+(v\tau/a)^2)+2\alpha\ln(\frac{\beta}{\pi\tau}\sinh(\pi\tau/\beta)),\label{eq:Q'}\\
Q''(\tau)&=2\alpha\arctan(v\tau/a).\label{eq:Q''}
\end{align}
These results together with Eq.~(\ref{eq:sigma_z}) allow us to write an expression for the current in the form
\begin{multline}
I(t)=\frac{\tilde q}{2}\Delta(t)\mathrm{Re}\int_0^t d \tau \Delta(\tau) e^{-Q'(t-\tau)-iQ''(t-\tau)}\\
\times [2n_a\cos[\Omega(\tau)-\Omega(t)]-e^{i(\Omega(\tau)-\Omega(t))}].
\label{eq:bosonization_current}
\end{multline}

	\subsection{Showing equivalence of solutions}
	\label{sec:Equivalence_of_solutions}
\begin{figure}[H]
\centering
\includegraphics[width=0.8\columnwidth]{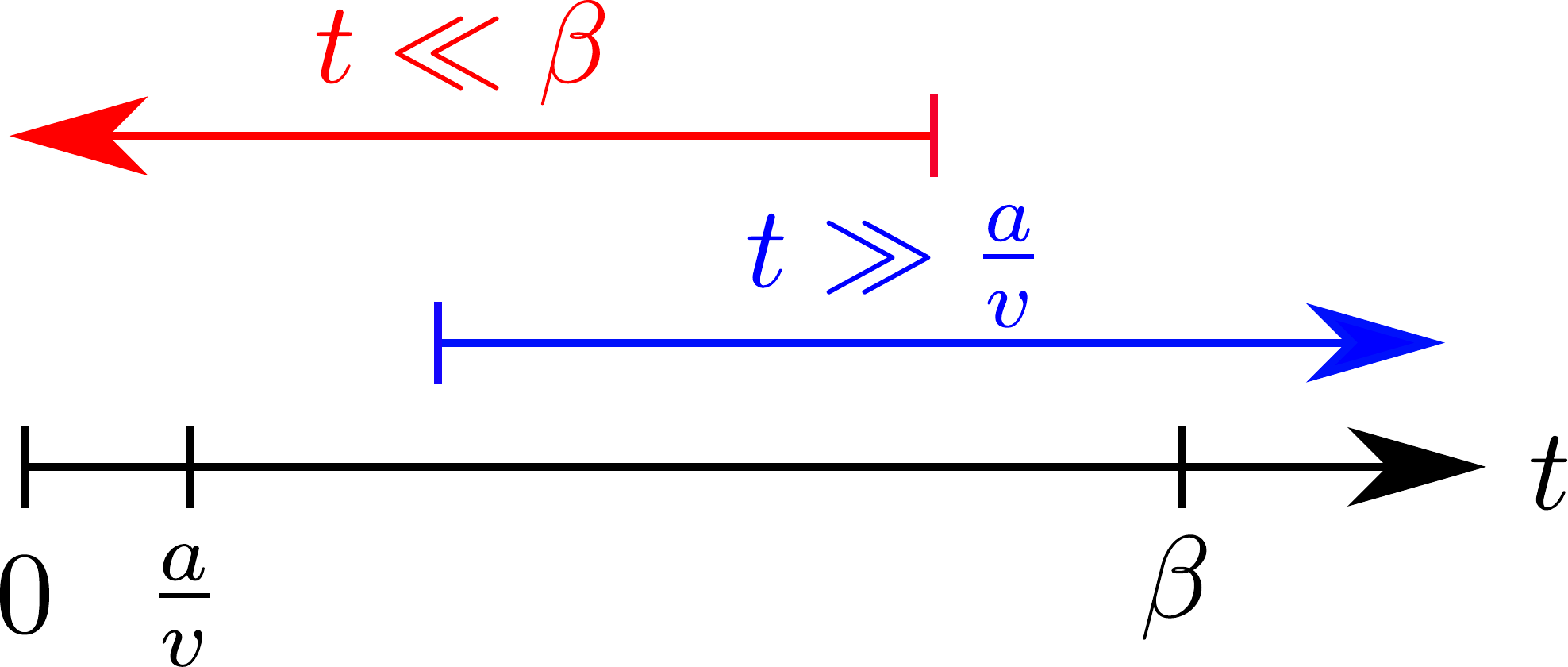}
\caption{Sketch showing that the whole of $t$-space can be divided into two regimes which overlap: the $t\gg\frac{a}{v}$ regime and the $t\ll\beta$ regime due to the fact that $\frac{a}{v}\ll\ll\beta$. Since we prove the identity \eqref{eq:equivalence_relation} in both limits, we have proven it for all $t$.}\label{fig:greater_than}
\end{figure}

	In this subsection, we will show that the current profile that was calculated using perturbation theory \eqref{current4} is equivalent to the result of the NIBA of the spin-boson model \eqref{eq:bosonization_current}. First, we want to show that $Q'$ and $Q''$ from section \ref{sec:Current_from_SB} are related to the propagator $\Phi$ from section \ref{sec:Current_from_bosonization} by 
	\begin{equation}
	\Phi(t)=\frac{1}{2\pi}a^{-\tilde\gamma^2}e^{-Q'(t)-iQ''(t)}.
	\label{eq:equivalence_relation}
	\end{equation}
	Since $\Phi^*(t)=\Phi(-t)$ and $Q'(t)=Q'(-t)$ and $Q''(t)=Q''(-t)$, we can focus on the case $t>0$. The important thing to realize is that $\frac{a}{v}\ll\ll\beta$, since $a$ is a small distance cut-off and the experimentally relevant regime is at low temperatures. Therefore the whole of $t$-space can be divided into two regimes which overlap: the $t\gg\frac{a}{v}$ regime and the $t\ll\beta$ regime, see Fig.~\ref{fig:greater_than}. From equation (\ref{eq:propagator}) and from (\ref{eq:Q'}) and (\ref{eq:Q''}) it is easy to show that Eq. (\ref{eq:equivalence_relation}) is satisfied in both limits. Hence we have proven their equality for all $t$. The identity \eqref{eq:equivalence_relation} can also be viewed as a consequence of the bosonization formalism, it is equivalent to Eq. (78) in Ref.~[\onlinecite{vonDelft1998}].
	
One can then show that \eqref{current4} and \eqref{eq:bosonization_current} are identical. We prove this in Appendix \ref{sec:equivalence}. We also prove that the current can be written in the more useful form
    \begin{multline}
	I(t)=\frac{\tilde q}{2}\Delta(t)\int_0^t\diff \tau\Delta(t-\tau)  \\\times \frac{\cos(2\alpha\arctan\frac{v\tau}{a}+\Omega(t-\tau)-\Omega(t))}{(1+(v\tau/a)^2)^\alpha\bigg(\frac{\beta}{\pi\tau}\sinh\frac{\pi\tau}{\beta}\bigg)^{2\alpha}}.
	\label{eq:electron_current}
	\end{multline}

	Since both solutions are entirely equivalent, when we refer to the perturbative solution in the text below, we are referring to either of the two solutions \eqref{current4} or \eqref{eq:bosonization_current}.
	
	\section{Non-perturbative approaches}
	
	\label{sec:non_perturbative}
	
	The two equivalent solutions we outlined above were both perturbative in the spin-bath coupling, however they were applicable for all $\alpha$. There are two special values of $\alpha$ for which we can go further and solve the problem to all orders in the coupling $\Delta$. In this section we briefly outline these two approaches and then show numerical data comparing the perturbative solution \eqref{eq:bosonization_current} to these exact methods. We show that at early times the perturbative solution gives very accurate results and can therefore be used to model the experimental set-up of Ref.~[\onlinecite{Feve2007}]. 
	
	The value $\alpha=1/2$ is special, because it corresponds to the case in which we have an integer quantum Hall edge \footnote{We did not consider the $\alpha=1/2$ case in our previous work \cite{OurLetter}, since that paper was focussed on the effect of Coulomb interactions on the current and these would renormalize $\alpha$ away from $1/2$ for an integer quantum Hall edge. Ref.~\cite{OurLetter} studied the generalized master equation which is valid for $\alpha\ll1$ and hence cannot be compared to the exact $\alpha=1/2$ solution. On the other hand, in the present work we study the perturbative approach that can be benchmarked again the exact solution.}. This means that we have a free fermion on the boundary and we can solve the problem exactly. This solution has been derived by previous authors \cite{Iwahori2016,Keeling2008}, however we present an alternative derivation in Appendix \ref{sec:IQH}.
	
	There is a further special point $\alpha=0$, in which case the quantum dot decouples completely from the edge and the problem becomes trivial. If we are close to this point, viz. $\alpha\ll 1$ then the entire perturbative expansion in $\Delta$ can be resummed as shown in [\onlinecite{Hartmann2000}]. The evolution of the dot is then given by the generalized master equation (GME). For more details of this approach see also Ref.~[\onlinecite{OurLetter}].

\begin{figure}[H]
		\includegraphics[width=9cm]{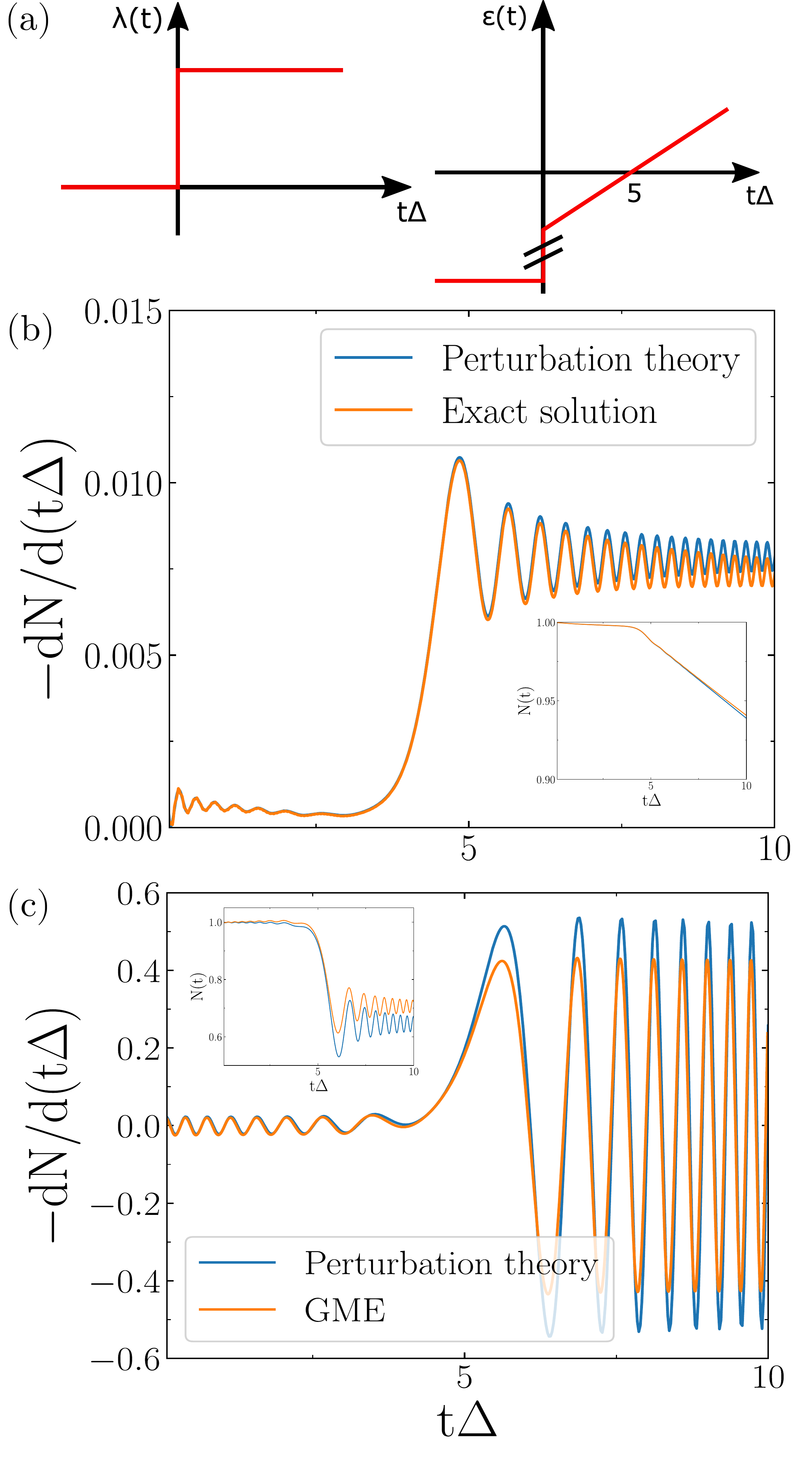}
		\caption{Comparison of the perturbative solution Eq. \eqref{current4} with two non-perturbative methods. We show the time evolution of the current after a linear ramp $\varepsilon(t)=\xi(t-t_0)$ with parameters $a=0.005v\Delta^{-1}$, $\xi=4\Delta^2$, $t_0=5\Delta^{-1}$. We plot $-dN/d(t\Delta)$ where $N(t)=\langle\hat N(t)\rangle$. The insets show the time evolution of the occupation number $N(t)$ on the QD. (a) Sketch of the sweep protocol. (b) For $\alpha=1/2$ we compare the exact solution as derived in Appendix \ref{sec:IQH} (yellow) with our perturbative result (blue). (c) For $\alpha=0.01$ we compare the result from the GME (yellow) with our perturbative result (blue). Detailed expressions for the GME are given in Ref. [\onlinecite{OurLetter}]. In both cases the early time agreement between the solutions is excellent.}%
		\label{fig:comparison}%
	\end{figure}

	We expect the perturbative solution to be valid at short times, for $t\Delta<1$. In Fig.~\ref{fig:comparison} we present our numerical results for the current and the occupation number on the QD after a linear voltage ramp with the rate $\xi$, so that $\varepsilon(t) = \xi (t-t_0)$. In this protocol, the dot is occupied in the initial state. At early times $\varepsilon(t)$ is negative and so only very little charge leaks off the dot as the dot equilibrates with the edge. For times  $t>t_0$ the bias becomes positive and the current greatly increases. The current shows oscillatory behaviour with increasing frequency as the bias increases with time. These are the characteristic Rabi oscillations.
	
	Fig.~\ref{fig:comparison}\textcolor{red}{(b)} shows the case $\alpha=1/2$, corresponding to a $\nu=1$ integer QHE state. We compare the exact solution \eqref{eq:IQHcurrent} to the perturbative result and find a good agreement at early times. At later times, the perturbative result misses the exponential decay of the current on the timescale $1/\Delta$. The inset shows that the occupation is initially very close to unity when $\varepsilon(t)<0$ and then starts decreasing once the $\varepsilon(t)>0$.
	Fig.~\ref{fig:comparison}\textcolor{red}{(c)} shows the result for $\alpha=0.01$, in which case the GME is expected to be a good approximation at all times. Our perturbative result agrees with the GME at early times as expected. For $t\Delta>5$ we start seeing a discrepancy between the two curves since the occupancy of the dot is starting to differ significantly from $1$ and the corresponding feedback effect leads to higher-order corrections to the current that our perturbation theory misses. Again, the inset shows the dot to be fully occupied until $\varepsilon(t)$ becomes positive. At late times the dot occupation tends to the Landau-Zener result \cite{OurLetter}. Comparing the $\alpha=0.01$ results to the $\alpha=1/2$ results, we see that the amplitude of the Rabi oscillations is strongly suppressed in the latter case. This is consistent with the crossover of the spin-boson model at $\alpha=1/2$ from the coherent to the incoherent regime. In the coherent regime there are strong oscillations, whereas in the incoherent regime the spin changes monotonically after a quench \cite{LeHur2009,LeHur2018}.
	
	Hence we see that in the two limits of $\alpha$ where we have access to simple solutions that are valid for all times, our perturbative result gives a good approximation to the current at early times. However, for other values of $\alpha$ which may be experimentally relevant no such simple solution schemes exist and here the perturbative solution \eqref{current4}, or equivalently the NIBA, is useful.

	\section{Analytical limits of the current profile}
	\label{sec:Analytical_limits}
	In this section, we show that we can obtain analytical expressions for the perturbative current profile \eqref{eq:bosonization_current} in a number of parameter regimes. We have three timescales\footnote{We always consider $\frac{a}{v}$ as our shortest time scale.} in our perturbation theory result: 
	\begin{equation}
	\quad \tau_B=\frac{\beta}{\pi},\quad \frac{1}{\Omega}=(\partial \ln\varepsilon/\partial t)^{-1}, \quad \frac{1}{\varepsilon_0}.
	\end{equation}
	$\frac{1}{\Omega}$ is the typical timescale on which the bias $\varepsilon(t)$ varies and $\varepsilon_0$ is the maximum amplitude of the bias. We assume the all the associated energy scales are much smaller than the FQH gap, which is of order $1\textrm{meV}$ for $\nu=1/3$.\cite{Kraphai2007}

	\subsection{Zero bias result}
In this set-up, a particle starts on the dot at zero bias $\varepsilon=0$ and the tunnelling is turned on suddenly at $t=0$ (but remains weak). The particle leaks slowly off onto the edge. 

	We consider general $\alpha=\frac{1}{2\nu}$ as appropriate for the case where an electron (as opposed to a quasi-electron) is tunneling. In the zero bias case $\varepsilon(t)=0$, although we only have the early time current using perturbation theory, we can extend the integration limit in (\ref{eq:bosonization_current}) to infinity, since the integrand vanishes quickly as $\tau\gg \frac{a}{v}$ \footnote{The term $\cos(2\alpha\arctan(v\tau/a))$ vanishes in this limit for $\alpha=\frac{1}{2\nu}=\frac{2n+1}{2}$}.
	This integral is solved in [\onlinecite{Leggett1987}] 
	\begin{equation}
	I_0=\frac{\tilde q}{\pi v}(2n_a-1)\tilde\lambda^2\frac{\sqrt{\pi}}{2}\frac{\Gamma(\alpha)}{\Gamma(\alpha+\frac{1}{2})}\bigg(\frac{\pi k_B T}{v}\bigg)^{2\alpha-1}
	\label{eq:zero_bias}
	\end{equation}
	\begin{equation}
	\frac{I_0}{\tilde q} \left\{\begin{array}{ll} >0  & n_a>1/2\\ =0 &  n_a=1/2 \\ <0 & n_a<1/2 \end{array} \right. 
	\end{equation}
	This makes sense from a physical point of view, the occupation number of the dot tends to $n_a=\frac{1}{2}$ as it reaches thermal equilibrium with the edge.
 
    We note that in Eq. \eqref{eq:zero_bias} the zero temperature limit is well-defined for $\alpha\geq1/2$. On the other hand, for $\alpha<1/2$ we require a finite temperature as an infrared cutoff.

	\subsection{Zero temperature, sinusoidal bias, $\alpha=\frac{3}{2}$}
\begin{figure}[H]
		\includegraphics[width=9cm]{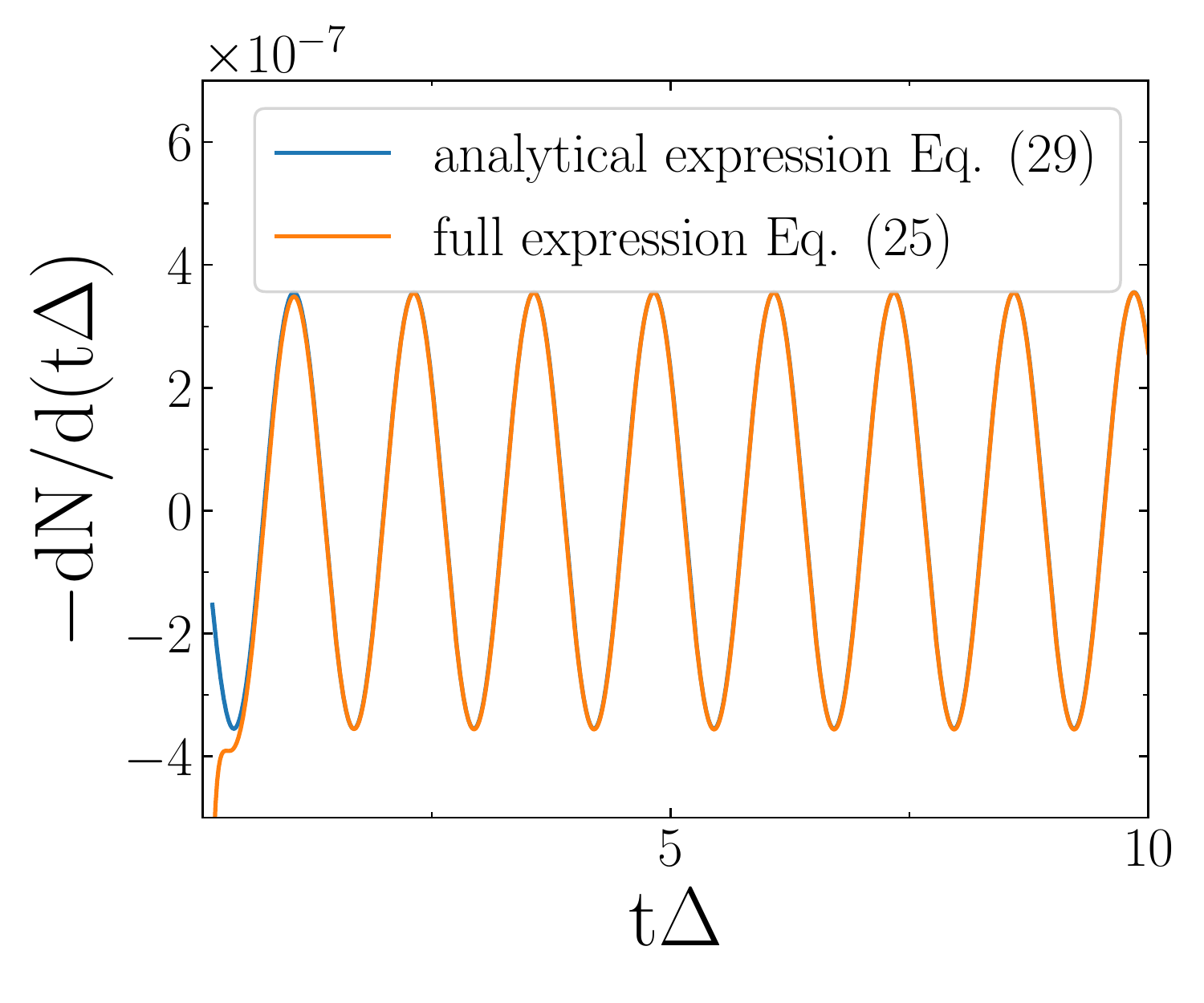}
		\caption{Comparison of the full solution Eq. \eqref{eq:electron_current} with the analytical expression Eq. \eqref{eq:analytic_current}, which is valid in the limit $\varepsilon_0\ll\Omega$. We show the time evolution of the current when the dot is driven with a bias $\varepsilon(t)=\varepsilon_0\cos\Omega t$. We use the parameters $\Omega=5\Delta$, $a=0.01v\Delta^{-1}$ and $\varepsilon_0=0.1\Delta$. The agreement between the solutions is excellent as long as $vt\gg a$.}%
		\label{fig:analytical}%
	\end{figure}

	Experiments must be performed at temperatures well below the FQH gap and therefore the zero temperature limit is the most relevant. Focusing on the $\nu=1/3$ particle case, when $\beta\to\infty$ the expression \eqref{eq:electron_current} simplifies.
	Let $\varepsilon(t)=\varepsilon_0\cos\Omega t$ and assume that $vt\gg a$ and $\varepsilon_0\ll\Omega$ so we can obtain the approximate form
	\begin{equation}
	I(t)=\frac{\tilde{q}\tilde\lambda^2}{2v^3}\frac{\Omega}{\pi} \varepsilon_0\bigg(\bigg[\ln \bigg(\frac{a\Omega}{v}\bigg)+\gamma_E\bigg]\sin\Omega t+\frac{\pi}{2}e^{a\Omega/v}\cos\Omega t\bigg),
	\label{eq:analytic_current}
	\end{equation}
	where $\gamma_E\approx0.577$ is the Euler-Mascheroni constant. We derive this result in detail in appendix \ref{sec:app_analytical}. We compare this analytical expression with the full integral expression \eqref{eq:electron_current} in the numerics presented in Fig.~\ref{fig:analytical} and see excellent agreement.  From this result we see that we will obtain a periodic current with a phase shift relative to the driving bias. We expect this phase shift to be an experimentally accessible signature.
	
	We note that there is a logarithmic dependence on the cut-off $a$ in the expression \eqref{eq:analytic_current}. This cut-off dependence is a generic feature for $\alpha>1/2$. Indeed, a similar behaviour is seen in numerical simulations of the spin-boson model using tensor network methods \footnote{G. Fux, J. Keeling and L. Brendon, Private communication.}. We can show explicitly that the results do not depend on the cut-off when $\alpha \leq 1/2$.
	
	\section{Conclusion}
	\label{sec:conclusion}
	In this paper, we derived in detail the relations between the quantum dot problem and the spin-boson model. The bosonized edge in the quantum dot problem maps to the bosonic heat bath of the spin-boson model. The quantum dot corresponds to the two-state system of the spin-boson model.
	We have perturbatively calculated the current arising when a quantum dot is coupled to an FQH edge. 
	We performed the calculations using the two alternative descriptions. To lowest order in perturbation theory in the spin-bath coupling $\lambda$, the solution obtained by bosonization of the original Hamiltonian is shown to be equivalent to the solution by mapping to the spin-boson model. This provides a consistency check of the map. 
	
	Our approach provides a very simple expression that can be used to compare to experimental results. We have shown that the perturbative calculation on the bosonized Hamiltonian agrees very well with two non-perturbative techniques. Firstly, the generalized master equation is a method for solving the spin-boson model when the coupling $\alpha$ is small. Secondly, we derive an exact solution of the non-interacting (IQH) problem. Numerical simulations show that as long as the occupation number on the dot stays close to its initial value, the agreement with our perturbative method is excellent. However, these exact methods are only valid in a limited parameter regime of the coupling $\alpha$. The perturbative solution is valid over the full parameter range of the spin-boson model, ie any $\alpha$. 
	
	For a periodic bias $\varepsilon(t)$ applied to the quantum dot, there is a phase shift between the bias $\varepsilon(t)$ and the resulting current $I(t)$. This theoretical prediction is verifiable experimentally. To perform the experiment, either the current on the edge after the dot can be measured directly or other indirect methods can be used. For example, one can couple a second quantum dot to the edge and drive it out of phase with the first dot in order to obtain zero current after the dot. An analogous experiment where the particle emitted by the quantum dot is reabsorbed by a quantum dot further along the channel was described for the IQH case in [\onlinecite{Moskalets2013}].
	
	Our perturbative calculation has also confirmed (see Appendix  \ref{sec:App_Klein_factors}) that neglecting the Klein factors in the bosonization prescription yields the correct answer for this model.
	
	We compared our prediction that the integrated charge in a current pulse is renormalized away from $-e$ due to Coulomb interactions to the experiment in Ref.~[\onlinecite{Feve2007}]. In particular, Fig. 1B of Ref.~[\onlinecite{Feve2007}] shows the time-dependent current on the edge. That experiment is performed for an integer quantum Hall edge, ie $\nu=1$ hence naively we would expect $\alpha=1/2$. However, as emphasized in Ref.~[\onlinecite{OurLetter}] and again in this paper, the Coulomb interactions between the dot and the edge will renormalize $\alpha$ to a different value. We searched for evidence of this interaction effect in the experimental data of Ref.~[\onlinecite{Feve2007}], however unfortunately the error bars on the experiment are too large, to detect the presumably small deviation from $\alpha=1/2$. We believe however, that the effect is large enough that it would be seen in experiments with improved accuracy. 
	
	Further theoretical work could be devoted to using the powerful numerical techniques---such as the stochastic Schr\"odinger equation---developed for the spin-boson model to model the quantum dot in experimentally relevant regimes. Recently, there has been a proposal to use tensor networks to study the spin-boson model \cite{Wall2016}. 
	
	Another possibility for further theoretical research would be to make use of the mapping to the Kondo problem to explore the Kondo regime of this problem more carefully. This problem should be tractable with DMRG techniques.
	\\
	\\
	\emph{Acknowledgements.}--- We are grateful to J. Keeling, G. Fux and L. Brendon for helpful discussions and for sharing results from their TEMPO algorithm. This work was supported by EP/N01930X/1 and EP/S020527/1. D.K. was supported by EPSRC Grant No. EP/M007928/2. Statement  of  compliance  with  EPSRC  policy  framework  on  research  data:  This  publication is theoretical work that does not require supporting research data.
	
	\appendix
	\section{Detailed perturbation theory calculations}
	\label{sec:perturb}
	In this appendix, we will derive the current profile \eqref{current4} using perturbation theory in detail. With the help of the spin commutation relations we can derive  explicitly  the current operator from the Heisenberg equation \eqref{eq:eomI}
	\begin{equation}
	\hat{I}(t)=-i\tilde q\tilde\lambda(t)\bigg(\hat{\tilde\psi}^{\dagger}(0,t)\hat S^-(t) - \hat S^+(t)\hat{\tilde\psi}(0,t)\bigg).
	\label{eq:current2}
	\end{equation}
	Combine equation \eqref{eq:current2} with the Kubo formula \eqref{eq:CurrentPer} we obtain the current expectation value at time $t$ as
	\begin{multline}
	\!\!\!\!\!I(t)=-\tilde q\tilde\lambda(t)\int_{0}^t\tilde\lambda(t')\left\langle [\hat S^-(t)\hat{\tilde\psi}^{\dagger}(0,t),\hat S^+(t')\hat{\tilde\psi}(0,t')]\right.\\ \left. -[\hat S^+(t)\hat{\tilde\psi}(0,t),\hat S^-(t')\hat{\tilde\psi}^{\dagger}(0,t') ]\right\rangle  \diff t'
	\label{eq:current3}
	\end{multline}
	In the interacting picture, the operators evolve with time under $H_0(t)$
	\begin{align}
	\hat S^-(t)=e^{-i\Omega(t)}\hat S^-(0),\quad \hat S^+(t)=e^{i\Omega(t)}\hat S^+(0),
	\label{eq:interactingpicture}
	\end{align}
	where $\Omega(t)=\int_0^t\varepsilon(s)\diff s$. Combining \eqref{eq:current3}, \eqref{eq:interactingpicture} and defining the fermionic propagators
	\begin{align}
	\Phi_{+-}(\tau)&=&\langle\hat{\tilde\psi}(0,\tau)\hat{\tilde\psi}^\dagger(0,0)\rangle,
	\label{eq:fermionic_propagator1}
	\\
	\Phi_{-+}(\tau)&=&\langle\hat{\tilde\psi}^\dagger(0,\tau)\hat{\tilde\psi}(0,0)\rangle,
	\label{eq:fermionic_propagator2}
	\end{align} 
	we obtain the result
    \begin{widetext}
	\begin{multline}
I(t)=-\tilde q\tilde\lambda(t)\int_{0}^t \diff t'\tilde\lambda(t')\bigg(e^{i\Omega(t')-i\Omega(t)}\bigg[(1-n_a)\Phi_{-+}(t-t')-n_a\Phi_{+-}(t'-t)\bigg] \\+e^{-i\Omega(t')+i\Omega(t)}\bigg[-n_a\Phi_{+-}(t-t')+(1-n_a)\Phi_{-+}(t'-t)\bigg] \bigg)
	\label{current44}
	\end{multline}
    \end{widetext}
using the time-translational invariance of the propagator.  Using the explicit form of fermion operator $\hat{\tilde\psi}$ in bosonization language \eqref{eq:fermion1} and the result for two points functions of vertex operators in [\onlinecite{vonDelft1998}]:
	\begin{equation}
	\langle e^{i\tilde\gamma\hat \varphi(\tau)}e^{-i\tilde \gamma\hat \varphi(0)}\rangle=\bigg(\frac{a}{v\tau_B\sin(\frac{iv\tau+a}{v\tau_B})}\bigg)^{\tilde\gamma^2},
	\end{equation}
	we can show that 
	\begin{equation}
	\Phi_{+-}(\tau)=\frac{1}{2\pi}\bigg(\frac{1}{iv\tau_B\sinh(\frac{v\tau-ia}{v\tau_B})}\bigg)^{\tilde\gamma^2}.
	\label{eq:propagator1}
	\end{equation}  
	One can also show that $\Phi_{+-}(\tau)=\Phi_{-+}(\tau)\equiv \Phi(\tau)$ \footnote{It can be derived easily from equation (87) of Ref. [\onlinecite{vonDelft1998}].}, which converts \eqref{current44} to the final result \eqref{current4}

	\section{Klein factors and (anti-)commutation relations} 
    \label{sec:App_Klein_factors}
	In the above calculation, we have neglected the Klein factors. The reason is that we only have once chiral edge and we are calculating the current. Klein factors become important when we have different species of particles. In perturbation theory, at all orders we have an equal number of $\hat \psi$ and $\hat \psi^\dagger$ in the expectation values and the Klein factors cancel. 
	
	\begin{equation}
	\hat{\tilde\psi}(x)=\frac{1}{\sqrt{2\pi}}a^{-\frac{\tilde\gamma^2}{2}}\hat F e^{-i\tilde\gamma\hat\varphi(x)}
	\label{eq:bos_with_KLein}
	\end{equation}
	where the Klein factors $\hat F$ satisfy \cite{vonDelft1998}
	\begin{equation}
	[\hat F,\hat b_k]=0,\ [\hat F, \mathcal{\hat N}]=\hat F \ \textrm{and}\ \hat F^\dagger\hat F=1,
	\end{equation}
where $\mathcal{\hat N}$ is the total number operator on the edge. Now if we substitute the expression \eqref{eq:bos_with_KLein} into \eqref{eq:current3}, then we can commute the Klein factors past the spin operators and use $\hat F^\dagger\hat F=1$ to eliminate the Klein factors.

We also note that if we replace the spin operators by ladder operators, viz. $\hat S^+=\hat a^\dagger$, then the Klein factors need to be commuted past both $\hat a$ and $\hat a^\dagger$ in \eqref{eq:current3} and so any statistical phase will cancel out.

\begin{widetext}
\section{Equivalence of two solutions: Details of the derivation}
\label{sec:equivalence}
We start off with the expression \eqref{current4}
\begin{multline}
I(t)=-\tilde q\int_{0}^t d t'\tilde\lambda(t)\tilde\lambda(t')\bigg(e^{i\Omega(t')-i\Omega(t)}[(1-n_a)\Phi(t-t')-n_a\Phi(t'-t)]+e^{i\Omega(t)-i\Omega(t')}[(1-n_a)\Phi(t'-t)-n_a\Phi(t-t')]\bigg)
\end{multline}
Substitute in the expression for $\Phi$ and use the fact that $Q'(t)=Q'(-t)$ and $Q''(t)=-Q''(-t)$ to find
\begin{equation}
I(t)=-\frac{\tilde q}{2\pi}a^{-\tilde\gamma^2}\int_{0}^t d t'\tilde\lambda(t)\tilde\lambda(t')e^{-Q'(t-t')}\bigg(e^{i\Omega(t')-i\Omega(t)}[(1-n_a)e^{-iQ''(t-t')}-n_ae^{iQ''(t-t')}]+c.c.\bigg),
\end{equation}
where $c.c.$ denotes the complex conjugate. We can sum the exponential to get and use $\Delta^2=\tilde\lambda^2\frac{2}{\pi}a^{-\tilde\gamma^2}$ to get
\begin{equation}
I(t)=-\frac{\tilde q}{2}\int_{0}^t d t'\Delta(t)\Delta(t')e^{-Q'(t-t')}\bigg([(1-n_a)\cos(\Omega(t')-\Omega(t)-Q''(t-t'))-n_a\cos(\Omega(t')-\Omega(t)+Q''(t-t'))\bigg)
\label{eq:intermediate_step}
\end{equation}
Going back to exponential notation
\begin{equation}
I(t)=-\frac{\tilde q}{2}\mathfrak{Re}\int_{}^t d t'\Delta(t)\Delta(t')e^{-Q'(t-t')-iQ''(t-t')}\bigg([1-n_a]e^{i\Omega(t')-i\Omega(t)}-n_ae^{-i\Omega(t')+i\Omega(t)}\bigg)
\end{equation}
\begin{equation}
I(t)=\frac{\tilde q}{2}\mathfrak{Re}\int_{}^t d t'\Delta(t)\Delta(t')e^{-Q'(t-t')-iQ''(t-t')}\bigg(2n_a\cos(\Omega(t')-\Omega(t))-e^{-i\Omega(t')+i\Omega(t)}\bigg)
\end{equation}
Finally, renaming $t'$ to $\tau$, we obtain the desired expression \eqref{eq:bosonization_current}. We also want to derive another useful expression. For $n_a=1$ we obtain from \eqref{eq:intermediate_step}
\begin{equation}
I(t)=\frac{\tilde q}{2}\int_{}^t d t'\Delta(t)\Delta(t')e^{-Q'(t-t')}\cos(\Omega(t')-\Omega(t)+Q''(t-t'))
\end{equation}
Defining $\tau=t-t'$ and substituting in the expressions \eqref{eq:Q'} and \eqref{eq:Q''} for $Q'$ and $Q''$ respectively, we find
    \begin{equation}
	I(t)=\frac{\tilde q}{2}\Delta(t)\int_0^t\diff \tau\Delta(t-\tau)  \frac{\cos(2\alpha\arctan\frac{v\tau}{a}+\Omega(t-\tau)-\Omega(t))}{(1+(v\tau/a)^2)^\alpha\bigg(\frac{\beta}{\pi\tau}\sinh\frac{\pi\tau}{\beta}\bigg)^{2\alpha}}.	
	\label{eq:electron_current_app}
	\end{equation}
and the zero bias result is
    \begin{equation}
	I_0(t)=\frac{\tilde q}{2}\Delta(t)\int_0^t\diff \tau\Delta(t-\tau)  \frac{\cos(2\alpha\arctan\frac{v\tau}{a})}{(1+(v\tau/a)^2)^\alpha\bigg(\frac{\beta}{\pi\tau}\sinh\frac{\pi\tau}{\beta}\bigg)^{2\alpha}}.	
	\end{equation}
For $\alpha=(2n+1)/2$ and $\Delta(t)=\theta(t)$, this $I_0(t)$ tends to a constant on the short timescale $a/v$
    \begin{equation}
	I_0=\frac{\tilde q}{2}\Delta^2\int_0^\infty\diff \tau \frac{\cos(2\alpha\arctan\frac{v\tau}{a})}{(1+(v\tau/a)^2)^\alpha\bigg(\frac{\beta}{\pi\tau}\sinh\frac{\pi\tau}{\beta}\bigg)^{2\alpha}}.	
	\end{equation}
Adding and subtracting $I_0$ from \eqref{eq:electron_current_app} we obtain 
	\small
    \begin{equation}
	I(t)=I_0+\frac{\tilde q}{2}\Delta(t)\int_0^t\diff \tau\Delta(t-\tau) \frac{1}{(1+(v\tau/a)^2)^\alpha}\frac{\cos(2\alpha\arctan\frac{v\tau}{a}+\Omega(t-\tau)-\Omega(t))-\cos(2\alpha\arctan\frac{v\tau}{a})}{\bigg(\frac{\beta}{\pi\tau}\sinh\frac{\pi\tau}{\beta}\bigg)^{2\alpha}}.	\label{eq:SB_result_app}
	\end{equation}
	\normalsize

\end{widetext}

	\section{Exact solution for IQH}
	\label{sec:IQH}
	This section presents an exact solution valid at the special point $\alpha=1/2$. The same result has been derived by previous authors using different methods, either by considering small time-slices over which the bias is constant \cite{Iwahori2016} or by calculating the S-matrix \cite{Keeling2008}. Here, we present an alternative method of calculating this result.
	
	In this section we represent the dot by fermionic creation and annihilation operators $\hat a^\dagger$ and $\hat a$. The transformation from fermions to spin-$1/2$ allows us to map between this representation and the spin representation used in the main text. In the integer case and in the absence of Coulomb interactions, the Hamiltonian is
	\begin{equation}
	\label{eq:Ham}
	\hat H=\hat H_0+\hat H_{\mathrm{tun}},
	\end{equation}
	where 
	\begin{align}
	\label{eq:freeHam}
	\hat H_0&=-i v\int_{-L/2}^{L/2}\hat \psi^\dagger(x)\partial_x\hat \psi(x)\diff x+\varepsilon(t)\hat{a}^\dagger(t) \hat{a}(t) 
	\end{align}
	is the free Hamiltonian. The first term of \eqref{eq:freeHam} describes the dynamics of an IQH edge with length $L$ at filling fraction $\nu=1$ \footnote{We explicitly set the chemical potential of the IQH edge to $\mu=0$.}. Wen showed in Ref. [\onlinecite{Wen1992}] that the edge modes of the IQH fluid are described by a free chiral fermion $\hat\psi$ whose velocity $v$ depends on the confining potential. The second term of \eqref{eq:freeHam} represents the quantum dot which we model as a time-dependent energy level  $\varepsilon(t)$. The coupling between the IQH edge and the quantum dot is modelled by the interaction term
	\begin{equation}
	\label{eq:HInt}
	\hat{H}_{\mathrm{tun}}(t)=\lambda(t)\hat\psi^\dagger(0) \hat a+h.c.
	\end{equation}
	To emphasize the position of the contact at $x=0$, we decompose the fermion field $\psi(x,t)$ into the left and right components
	\begin{equation}
	\label{eq:split}
	\hat \psi(x,t)=\hat \psi_L(x,t)\Theta(-x)+\hat \psi_R(x,t)\Theta(x),
	\end{equation}
	where $\Theta(x)$ is the Heaviside step function and we use the convention $\Theta(0)=1/2$ to symmetrize the contribution of the left and right parts at the contact point. From the Hamiltonian \eqref{eq:Ham}, we derive the field equations
	\begin{align}
	\!\!\!\!\!\! i\partial_t\hat{a}(t)&=\frac{\lambda(t)}{2}(\hat \psi_L(0,t)+\hat \psi_R(0,t))+\varepsilon(t) \hat a(t),\label{eq:EL1}\\
	\!\!\!\!\!\!i\dot{\hat \psi}(y,t)&=-i v\partial_y\hat \psi(y,t)+\lambda(t) \delta(y)\hat a(t),
	\label{eq:EL11}
	\end{align}
	where we have already used the decomposition \eqref{eq:split}. Integrating \eqref{eq:EL11} from $-\epsilon$ to $+\epsilon$, we arrive at the constraint \footnote{The LHS is $\mathcal{O}(\epsilon)$ and thus can be neglected.}
	\begin{equation}
	0=i v (\hat \psi_R(0,t)-\hat \psi_L(0,t))-\lambda(t)  \hat a(t).
	\label{eq:EL2}
	\end{equation}
	With the help of \eqref{eq:EL2}, we can eliminate $\hat\psi_R$ from \eqref{eq:EL1} and arrive at 
	\begin{equation}
	\!\!\!\!\!\!\! i\partial_t \hat a(t)=\frac{\lambda(t)}{2}\bigg(2\hat \psi_L(0,t)-i\frac{\lambda(t)}{v}\hat a(t)\bigg)+\varepsilon(t)\hat a(t).
	\label{eq:eom}
	\end{equation}
	Since the IQH edge is described by a chiral fermion $\hat\psi$, the appearance of the quantum dot only affects the right component of the IQH edge \footnote{The situation is much more complicated if the fermion $\hat\psi$ is neither chiral nor free.}. With this observation, we can expand $\hat\psi_L(x,t)$ in terms of free modes
	\begin{align}
	\hat \psi_L(x,t)=\frac{\sqrt{L}}{v}\int_{-\infty}^{\infty} \frac{\diff \omega}{2\pi} e^{i\omega(\frac{x}{v}- t)}\hat c_\omega,
	\label{eq:modes}
	\end{align}
	where the fermion operator $\hat c_\omega$ annihilates a chiral mode at energy $\omega$ on the IQH edge. Substituting \eqref{eq:modes} into \eqref{eq:eom} and using the ansatz 
	\begin{equation}
	\label{eq:IQH_exact}
	\hat a(t)=g(t)\hat a(0)+\int_{-\infty}^{\infty} \frac{\diff \omega}{2\pi} f_\omega(t)\hat c_\omega,
	\end{equation}
	we obtain differential equations for $g(t)$ and $f_{\omega}(t)$. Solving these differential equations, we derive the exact solution
	\begin{align}
	\!\!\!\!\!\!\!g(t)&=e^{-\xi(t)},\label{eq:g}\\
	\!\!\!\!\!\!\!\!\!\!\!\!\!f_\omega(t)&=-i\frac{\sqrt{L}}{v}\int_{}^t\diff t' \lambda(t')e^{-i\omega t'+\xi(t')-\xi(t)},\label{eq:f}
	\	\end{align}
	where we have defined
	\begin{equation}
	\xi(t)=\int_{-\infty}^{t}\bigg[i\varepsilon(s)+\frac{\lambda(s)^2}{2v}\bigg]\diff s
	\end{equation}
	In the next section we set $\lambda(t)=\lambda\Theta(t)$ for conciseness. In that case we define $\Omega(t)=\int_0^t\varepsilon(t)\diff t$ and the timescale 
	\begin{equation}
	\tau_0=\frac{v}{\lambda^2}
	\end{equation}
	which is the timescale over which a current from the dot to the edge decays and it can hence be viewed as a tunnelling timescale. With this result, we are able to derive the quantities that can be measured in the physical set-up with the general applied bias voltage $\varepsilon(t)$.

	We define the current operator via \eqref{eq:eomI}. From the exact time-dependent operator $\hat{a}$ \eqref{eq:IQH_exact}, we can derive the expectation value of the current at any given time $I(t)=\langle\hat I (t)\rangle$ using the Heisenberg picture. In order to calculate the expectation value of the time-dependent operator, we need to set the initial condition of the quantum dot and introduce the Fermi distribution on the IQH edge
	\begin{align}
	\langle \hat{a}^\dagger(0) \hat{a}(0)\rangle&=n_a, \label{eq:initial}\\
	\langle \hat{c}_{\omega'}^\dagger \hat{c}_\omega\rangle&=\frac{2\pi v}{L}\delta(\omega-\omega')n_F(\omega),\label{eq:Fermi}
	\end{align}
	where $n_F(\omega)=(e^{\beta \omega}+1)^{-1}$ is the Fermi distribution and we define $\beta=1/k_B T$ as usual. In order to obtain sensible results, we need to introduce a cutoff frequency $\omega_c \gg k_BT$, which adds a factor $e^{\omega/\omega_c}$ to the frequency integrals. This cut off makes sense physically since the negative frequency fermion modes, which are deep inside the Fermi sea, do not affect the low energy physics near the chemical potential. We write this in the suggestive form $\omega_c=v/a$. Combining the exact solution \eqref{eq:IQH_exact} and the definition \eqref{eq:eomI}, after some algebraic calculations, we obtain   
	\begin{widetext}
		\begin{multline}
		\label{eq:IQHcurrent}
		I(t)=q\frac{\lambda(t)^2}{v}\mathfrak{Re}\bigg\{e^{-2\mathfrak{Re}\xi(t)}n_a-\frac{e^{-2\mathfrak{Re}\xi(t)}}{v}\int_{-\infty}^t\diff t'\int_{-\infty}^t\diff t''\lambda(t')\lambda(t'')\frac{1}{2\pi}\frac{\pi i}{\beta}\frac{e^{\xi(t')+\xi^*(t'')}}{\sinh\bigg(\frac{\pi}{\beta}(t''-t'-\frac{ia}{v})\bigg)}\\
		+2\int_{-\infty}^t\diff t'\frac{1}{2\pi}\frac{\pi i}{\beta}\frac{e^{\xi(t')-\xi(t)}}{\sinh\bigg(\frac{\pi}{\beta}(t-t'-\frac{ia}{v})\bigg)}
		\bigg\}.
		\end{multline}
	\end{widetext}
	We can show explicitly, that our space cutting solution obeys charge conservation. We have the operator identity\footnote{In this section, we omit all the $(x,t)$ arguments for simplicity.}  
	\begin{equation}
	i\partial_t(\hat a^\dagger a)=i\hat a^\dagger\partial_t\hat a+i(\partial_t\hat a^\dagger)\hat a.
	\end{equation}
	Combining above equation with (\ref{eq:EL1}) and its complex conjugate, we obtain
	\begin{equation}
	i\partial_t(\hat a^\dagger a)=\frac{\lambda}{2}\bigg(\hat a^\dagger (\hat \psi_R+\hat \psi_L)-(\hat \psi_R^\dagger+\hat \psi_L^\dagger)a\bigg).
	\end{equation}
	Now replacing $\hat a$ and $\hat a^\dagger$ using (\ref{eq:EL2}) we find
	\begin{equation}
	\label{eq:conserved}
	\partial_t(\hat a^\dagger a)=v(\hat \psi_L^\dagger\hat \psi_L-\hat \psi_R^\dagger\hat \psi_R).
	\end{equation}
	which is nothing but the charge conservation equation. The left hand side of \eqref{eq:conserved} is the time variation of total charge on the quantum dot. The right hand side of \eqref{eq:conserved} is the total current from the IQH edge to the dot since the current that goes into the contact point is $v(\hat \psi_L^\dagger\hat \psi_L)$ and the current that goes out of it is $v(\hat \psi_R^\dagger\hat \psi_R)$. If we add the Coulomb interaction term $H_\textrm{int}\propto\hat \psi^\dagger(0)\hat \psi(0) \hat a^\dagger \hat a$ to the Hamiltonian, then the equation of motion \eqref{eq:EL11} is modified. However, it is easy to show that even then, the charge conservation equation \eqref{eq:conserved} is still satisfied.
    
We now compare the exact solution to our perturbative result (\ref{current4}). Setting $\tilde\gamma=1$ in (\ref{eq:propagator}) and using the assumption $v\tau_B/a\to\infty$ we obtain
	$\mathfrak{Re}\Phi(t-t')=\frac{1}{2v}\delta(t-t')$. Setting $\lambda(t)=\lambda\Theta(t)$, we find the final result for the case $\tilde\gamma=1$
	\begin{multline}
	\!\!\!\!\!\!\!\!\!\! I(t)=\frac{qn_a\lambda^2}{v}+\frac{q\lambda^2}{\pi v}\mathfrak{Re}\bigg[\int_0^{t} \diff t'\quad e^{i\Omega(t')-i\Omega(t)}\\ \frac{\pi i}{\beta}\frac{1}{\sinh\bigg(\frac{\pi (t-t')}{\beta}-i\frac{\pi a}{\beta v}\bigg)}\bigg],
	\label{eq:IQH_check}
	\end{multline}
	where we have used $\Theta(0)=1/2$. If we expand the exact solution \eqref{eq:IQHcurrent} to second order in $\lambda$, then equation \eqref{eq:IQH_check} is obtained, so indeed both methods agree. This equivalence is a non-trivial cross check of our perturbation theory method. 
	
	\section{Derivation of the analytical expression for the current}
	\label{sec:app_analytical}
For conciseness define the frequency scale $\omega_c=v/a$. When $\beta\to\infty$ the expression \eqref{eq:electron_current} simplifies to
\begin{equation}
I(t)=\frac{e\Delta^2}{4}\int_0^t \diff \tau\frac{\cos(3\arctan(\omega_c\tau)+\Omega(t-\tau)-\Omega(t))}{(1+(\omega_c\tau)^2)^{3/2}}
\end{equation}
Define
\begin{equation}
\zeta(t-\tau,t)=\Omega(t-\tau)-\Omega(t)
\end{equation}
Now assume that $\epsilon_0\ll\Omega$ so $\zeta\ll1$ and we can expand the trigonometric functions:
\begin{multline}
I(t)=\frac{e\Delta^2}{4}\int_0^t \frac{\diff \tau}{(1+(\omega_c\tau)^2)^{3/2}}\\\times\bigg(\cos(3\arctan(\omega_c\tau))\cos\zeta(t-\tau,t)\\-\sin(3\arctan(\omega_c\tau))\sin\zeta(t-\tau,t)\bigg)
\end{multline}
\begin{multline}
I(t)\approx\frac{e\Delta^2}{4}\int_0^t \frac{\diff \tau}{(1+(\omega_c\tau)^2)^{3/2}}\\\times\bigg(\cos(3\arctan(\omega_c\tau))-\sin(3\arctan(\omega_c\tau))\zeta(t-\tau,t)\bigg)
\end{multline}
Now use the trigonometric identities
\begin{equation}
\cos(3\arctan x)=\frac{1-3x^2}{(1+x^2)^{3/2}}
\end{equation}
\begin{equation}
\sin(3\arctan x)=\frac{3x-x^3}{(1+x^2)^{3/2}}
\end{equation}
so with $\epsilon(t)=\epsilon_0\cos\Omega t$ 
\begin{multline}
I(t)\approx\frac{e\Delta^2}{4}\int_0^t \frac{\diff \tau}{(1+(\omega_c\tau)^2)^{3}}\bigg[1-3(\omega_c\tau)^2\\-\frac{\epsilon_0}{\Omega}(3\omega_c\tau-(\omega_c\tau)^3)\bigg(\sin\Omega t(1-\cos\Omega\tau)+\cos\Omega t\sin\Omega\tau\bigg)\bigg]
\end{multline}
and rescaling the integration variable
\begin{multline}
I(t)\approx\frac{e\Delta^2}{4\omega_c}\int_0^{\omega_ct} \frac{\diff x}{(1+x^2)^{3}}\bigg[1-3x^2\\-\frac{\epsilon_0}{\Omega}(3x-x^3)\bigg(\sin\Omega t(1-\cos\frac{\Omega}{\omega_c}x)+\cos\Omega t\sin\frac{\Omega}{\omega_c}x\bigg)\bigg]
\end{multline}
and
\begin{equation}
I(t)=I_0(t)+I_1(t)\sin\Omega t+I_2(t)\cos\Omega t
\label{eq:R1}
\end{equation}
where
\begin{multline}
I_0(t)=\frac{e\Delta^2}{4\omega_c}\int_0^{\omega_ct} \diff x\frac{1-3x^2}{(1+x^2)^{3}}=\frac{e\Delta^2}{4\omega_c}\frac{\omega_c t}{(1+(\omega_ct)^2)^2}\\\approx \frac{\Delta^2}{4\omega_c}\frac{1}{(\omega_ct)^3}\to 0
\end{multline}
as $\omega_c t\to\infty$.
Now
\begin{equation}
I_1(t)=\frac{e\Delta^2}{4\omega_c}\frac{\epsilon_0}{\Omega}\int_0^{\omega_ct} \diff x\frac{(3x-x^3)(1-\cos\frac{\Omega}{\omega_c}x)}{(1+x^2)^{3}}
\end{equation}
Since $\omega_c t\gg1$ and the integrand dies off at large $x$, we can extend the upper limit of the integral to $\infty$. Then this integral can done exactly in terms of the Meijer G-function. We can expand this for $\frac{\Omega}{\omega_c}\ll 1$

\begin{equation}
I_1(t)=\frac{e\Delta^2}{4\omega_c}\frac{\epsilon_0}{\Omega}\frac{1}{4}  \left(2 \ln \frac{\Omega}{\omega_c}+2\gamma_E\right)\bigg(\frac{\Omega}{\omega_c}\bigg)^2+\mathcal{O}\bigg(\frac{\Omega}{\omega_c}\bigg)^3
\label{eq:R2}
\end{equation}

Now 
\begin{equation}
I_2(t)=\frac{e\Delta^2}{4\omega_c}\frac{\epsilon_0}{\Omega}\int_0^{\omega_ct} \diff x\frac{(3x-x^3)\sin\frac{\Omega}{\omega_c}x}{(1+x^2)^{3}}
\end{equation}
As before extend the upper limit to infinity
\begin{equation}
I_2(t)=\frac{e\Delta^2}{4\omega_c}\frac{\epsilon_0}{\Omega}\int_0^{\infty} \diff x\frac{(3x-x^3)\sin\frac{\Omega}{\omega_c}x}{(1+x^2)^{3}}
\end{equation}
and expanding this:
\begin{equation}
I_2(t)=\frac{e\Delta^2}{4\omega_c}\frac{\epsilon_0}{\Omega}\frac{\pi}{4}e^{-\Omega/\omega_c}\bigg[\bigg(\frac{\Omega}{\omega_c}\bigg)^2+\mathcal{O}\bigg(\frac{\Omega}{\omega_c}\bigg)^4\bigg]
\label{eq:R3}
\end{equation}
Substituting \eqref{eq:R2} and \eqref{eq:R3} into \eqref{eq:R1}, we obtain \eqref{eq:analytic_current}.
	

	\bibliography{bib.bib}

\end{document}